\documentclass[twocolumn,showpacs]{revtex4}
\usepackage{epsfig}
\begin{document}
\title{Experimental observation of Loschmidt time reversal of a Quantum
Chaotic System}
\author{A. Ullah and M.D. Hoogerland}
\affiliation{Department of Physics, University of Auckland, Private
Bag 92019, Auckland, New Zealand}

\newcommand{\kbar}{\mathchar'26\mkern-9muk}

\begin{abstract}
We have performed an experiment to demonstrate the approximate time reversal of a ``chaotic'' time evolution of atomic deBroglie waves. We use ultra cold atoms from a Bose-Einstein condensate in
a quantum $\delta$-kicked rotor experiment, and show that an initial state can be approximately re-created even after a period of ``chaotic'' evolution (a number of kicks). As this mechanism
only works for a very narrow range of momenta, the net effect is a narrowing of the momentum distribution after the kick sequence.
\end{abstract}

\maketitle


How the irreversibility of macroscopic systems is reconciled with
the reversibility of microscopic physical laws has been discussed
for over a century. The argument was first made by J. Loschmidt
\cite{Loschmidt} in reaction to Boltzmann's statistical theory of
gases. These were the ideas which helped to develop the theory of
dynamical chaos \cite{Chirikov1,kornfeld}. Small perturbations may
grow exponentially with time in classical dynamics which makes the
motion practically irreversible. This phenomenon is known as
classical chaos \cite{Lieberman1,Lieberman2}. In quantum dynamics on
the other hand, chaos does not exist. Exponential divergence takes
place only during the rather short Ehrenfest time \cite{chirikov2},
and the quantum evolution remains stable and reversible in the
presence of small perturbations.

The delta-kicked rotor is a simple system well suited to the study
of classical and quantum chaos
\cite{Raizen,Robinson,Shepelyansky,Haake}. As a consequence of the
quantum-mechanical nature of the motion, dynamical localization has
been observed in this system \cite{Ammann,Moore}. It has been
central to the study of ``quantum chaos'', receiving significant
attention in recent years \cite{saif,Currivan}. Casati and
co-workers predicted that a quantum particle follows the diffusive
dynamics of a classical chaotic system only up to a certain time,
known as the quantum break time \cite{casatiandprozen,casati}. The
classical diffusion ceases due to quantum interference, and the
momentum distribution settles into an exponential distribution at
this point.

Here we present an experimental realization of the effective time
reversal of atomic matter waves as proposed by Martin {\em et. al.}
\cite{Martin}. We use a cloud of ultra cold $^{87}$Rb atoms, driven
by a pulsed optical standing wave, as a quantum $\delta$-kicked
rotor system. We show that the effects of a certain number of kicks
can be effectively reversed by further kicks by having a different
kick sequence. We observe that a significant fraction of the atoms
return back to their original zero momentum state. As the time
reversal is very sensitive to the original momentum, we observe a
narrowing of the zero momentum peak. This phenomenon has been named
``Loschmidt cooling'' \cite{Martin}, even though the phase space
density is not increased. It should be stressed that many schemes
exist that narrow the momentum distribution of a sample of atoms
\cite{Gong,cooling,Nelson}, some of which increase the phase space
density. The results presented here however, show the first
observation of an effective reversal of chaotic dynamics in the
quantum regime. The phenomenon observed here can be thought of as a
multiple beam interferometry, where the sharpness of the
interference fringes due to constructive interference has been
demonstrated \cite{Weitz}.

The atom optics implementation of the delta-kicked rotor consists of
a two-level atom placed in a pulsed standing wave of laser light
that is detuned from resonance. The laser field gives rise to a
potential that varies sinusoidally with position. The Hamiltonian of
the system can be written as
\begin{equation}
\ H =\frac{p^2}{2m}-V_0\cos (k_L x) f(t),
\end{equation}
where $f(t)$ describes the time dependence of the laser pulses. For
the delta kicked rotor, $f(t)=\sum_{n=1}^N\delta(t-nT)$, where $T$
is the kick period. In this system, it is convenient to use the
scaled kick period $\kbar=8\omega_RT$ where $\omega_R=\hbar
k_L^2/(2m)$ is the recoil frequency, with $k_L=2\pi/\lambda$,
$\lambda$ is the wavelength of the laser beam and $m$ the mass of
the atom. A ``quantum resonance'' exists for $\kbar=4\pi$, where all
kicks add coherently, and quadratic energy growth with the number of
kicks is observed. At $\kbar=2\pi$, an ``anti-resonance'' is
observed, where the effect of each kick is effectively negated by
the following kick.

The potential $V_0$ can be written as $\hbar \phi_d$, where
$\phi_d=\tau_p\Omega^2/4\Delta$ is the kick strength, $\tau_{p}$ is
the pulse duration, and $\Omega=d \cdot E/(2\hbar)$ is the
on-resonance Rabi frequency. The parameter $d$ is the atomic dipole
moment induced by the laser with electric field $E$.
$\Delta=\omega_L-\omega_0$ is the detuning from resonance, where
$\omega_L$ is the laser frequency and $\omega_0$ is the resonance
frequency.

Following \cite{Martin} we use a Loschmidt pulse train consisting of
 $N/2$ pulses with a scaled period of $\kbar=4\pi+\epsilon$ , a
 waiting time $6\pi$ after the last pulse and then $N/2$ pulses with
 a period of $4\pi-\epsilon$, thus a total of $N$ pulses.
  The parameter $\epsilon$ is proportional to the difference of the
pulse period from the first primary resonance.

We simulate the time evolution of the atom optics kicked rotor using
the split operator method as described in \cite{maarten}. We Fourier
transform our wave function from position space into momentum space
and back again, as the kick potential is a diagonal operator in
position space, and the free evolution is a diagonal operator in
momentum space. The kicks are sufficiently short that the evolution
due to the momentum can be ignored during the kicks, which is
equivalent to the Raman-Nath regime. We start the simulation with a
Gaussian wave packet with a width (2$\sigma$) of 0.1 recoils in
momentum space.

\begin{figure}
\includegraphics[width=\columnwidth]{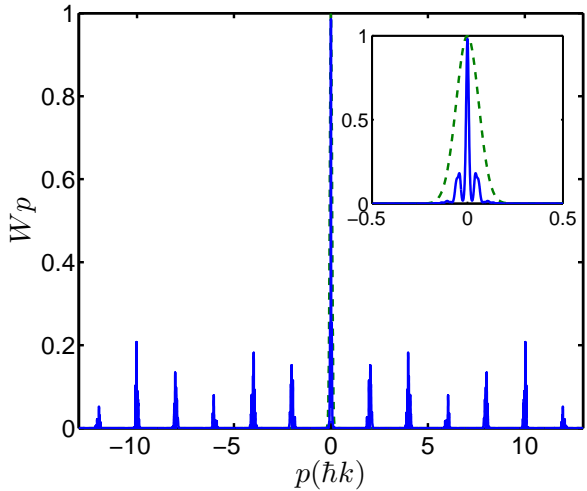}
\caption{\label{fig:returnpeak} (color online) The simulated
momentum distribution after the kick sequence (full blue line) and
the initial wave packet (dashed green line) are shown. The
parameters are $N=10$, $\phi_{d}\sim 2.5$ and $\epsilon=2$. The
inset shows a magnified view of the area around $p=0$.}
\end{figure}

The momentum distribution after the kick sequence can be written as
$W_p(t)=|\langle p|\psi(t)\rangle|^2/|\langle 0|\psi(0)\rangle|^2$.
In Fig. \ref{fig:returnpeak} we show the resulting momentum
distribution in the simulation after a pulse sequence as discussed.
A much narrowed momentum distribution around $p=0$ after the kick
sequence is observed, as the restitution of the wave function only
works for very small values of $p$. The rest of the probability is
transferred to higher momentum states, with offsets of $2n\hbar k$
with integer $n$. Note however, that the probability for $p=0$
returns to its previous value. Martin {\em et al.} \cite{Martin}
have performed extensive simulations, showing that the $p=0$ peak
gets narrower as the number of kicks $N$ in the sequence or the
intensity increases. This does however mean that fewer atoms are in
this peak than at lower intensities and/or lower kick number. Note
that the final momentum distribution is not Gaussian. We therefore
take the FWHM as the indicator for the width of the peak.

\begin{figure}
\includegraphics[width=\columnwidth]{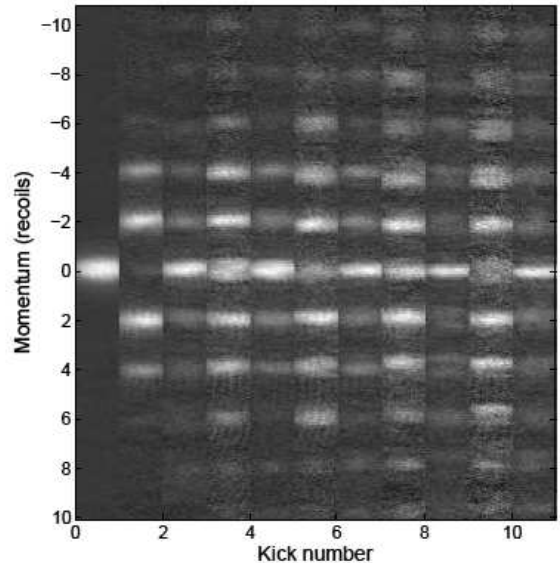}
\includegraphics[width=\columnwidth]{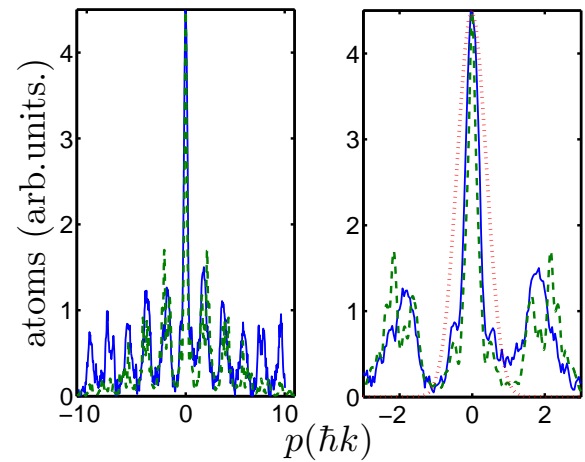}
\caption{\label{fig:lastpeak} (color online) (top) Absorption images of the momentum distribution before and after each kick (1-10). (bottom) The momentum distribution of the atoms at $N$ kicks
(bottom left) for the experiment (solid blue) and the simulation (dashed green). The central part of the same curve is also shown (bottom right), with the initial state in (dotted red). The
parameters are $N=10$, $\phi_{d}\approx 2$ and $\epsilon=1$.}
\end{figure}

For the experiment, a reasonable number of atoms at small initial
momenta is of paramount importance, and we therefore use a
Bose-Einstein condensate (BEC) of ($F=1$) $^{87}$Rb at a temperature
of 50~nK as the source of atoms. The atoms are first captured and
cooled in a magneto-optical trap (MOT) then transferred using a push
laser to a second MOT in a second, connected chamber. An ultra-high
vacuum is produced in the second chamber by an ion pump connected to
it.  The condensate is formed in a dipole trap overlapped with the
second MOT. The dipole trap is made by a pair of intersecting
focused CO$_2$ laser beams. A detailed description of the
experimental setup can be found in \cite{maarten}.

We realize the AOKR by pulsing a near resonant optical standing
wave, derived from a 780 nm diode laser onto a BEC of $\sim 10^4$
$^{87}$Rb atoms. The trap containing the BEC is turned off 500
$\mu$s before the kick sequence to reduce mean field effects. The
kick laser is locked to the $S_{\frac{1}{2}}, F=2\rightarrow
P_{\frac{3}{2}}, F'=3$ transition in $^{85}$Rb isotope. Hence, the
laser frequency is detuned by 2.45~GHz from the relevant
$F=1\rightarrow F=2$ resonance frequency. The laser beam from the
diode laser passes through a $50/50$ beam splitter and the output
beams are then passed through separate acousto-optic modulators
(AOMs) for fast switching. After passing through the AOMs, the two
beams pass through single mode optical fibres and are focussed onto
the BEC from opposite directions to produce a standing wave. The
beam diameter at the focus is $\sim 100\mu$m.

The laser pulses of the standing wave are the kicks that modify the
momentum distribution of the atoms. The momentum distribution of the
atoms after the kick sequence is measured by absorption imaging in
time of flight, with a flight time of typically 8~ms. Just prior to
imaging, the atoms are optically pumped to the $F=2$ state by a
100~$\mu$s pulse on the $F=1\rightarrow F'=2$ repump transition. An
absorption image is then obtained using a probe laser which is tuned
to the $S_{\frac{1}{2}}, F=2\rightarrow P_{\frac{3}{2}}, F'=3$
transition. Typical results are shown in Fig. \ref{fig:lastpeak}
(Top), where we show the absorption images of the momentum
distribution before kicking, and the distribution after each kick.
We sum the images along the rows to obtain a one-dimensional
momentum distribution, which can be compared to the simulation. The
resulting momentum distribution of the ensemble after $10$ kicks
with $\epsilon=1$ (bottom left) and the magnified view of central
part of the same distribution are also shown in the figure. Note
that some of the structure in the final distribution of
Fig.~\ref{fig:returnpeak} can be observed as a pedestal on the final
zero momentum peak in the experiment.

After deconvolution with the original trap size, the width of the
zero velocity peak after 10 kicks is $\sigma =0.21$ (recoils) as
compared to $\sigma =0.43$ (recoils) for the initial momentum
distribution without kicking. It should be noted that the height of
the zero momentum peak after the full sequence does not return to
the height of the original BEC peak, which is currently due to
experimental limitations, mainly the resolution of the imaging
system and the observable amount of absorption.

The Loschmidt time reversal works for a very narrow range of initial
momenta, which depends on the value of $\epsilon$, the kick
strength, and the total number of kicks. The more narrow this range,
the fewer atoms are part of it. Hence, experimental limitations make
it difficult to observe extremely narrow momentum distributions.

\begin{figure}
 \includegraphics[width=\columnwidth]{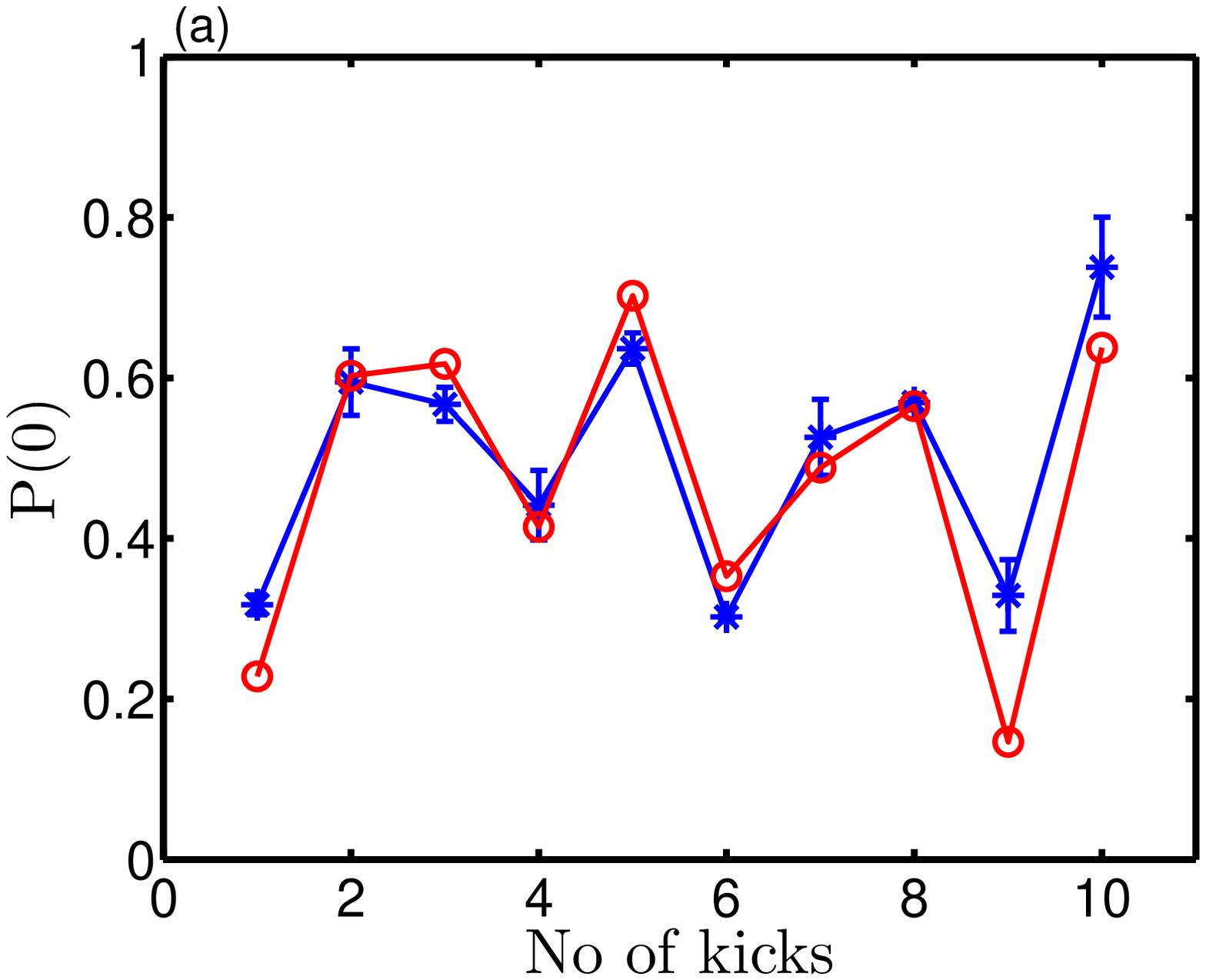}
 \includegraphics[width=\columnwidth]{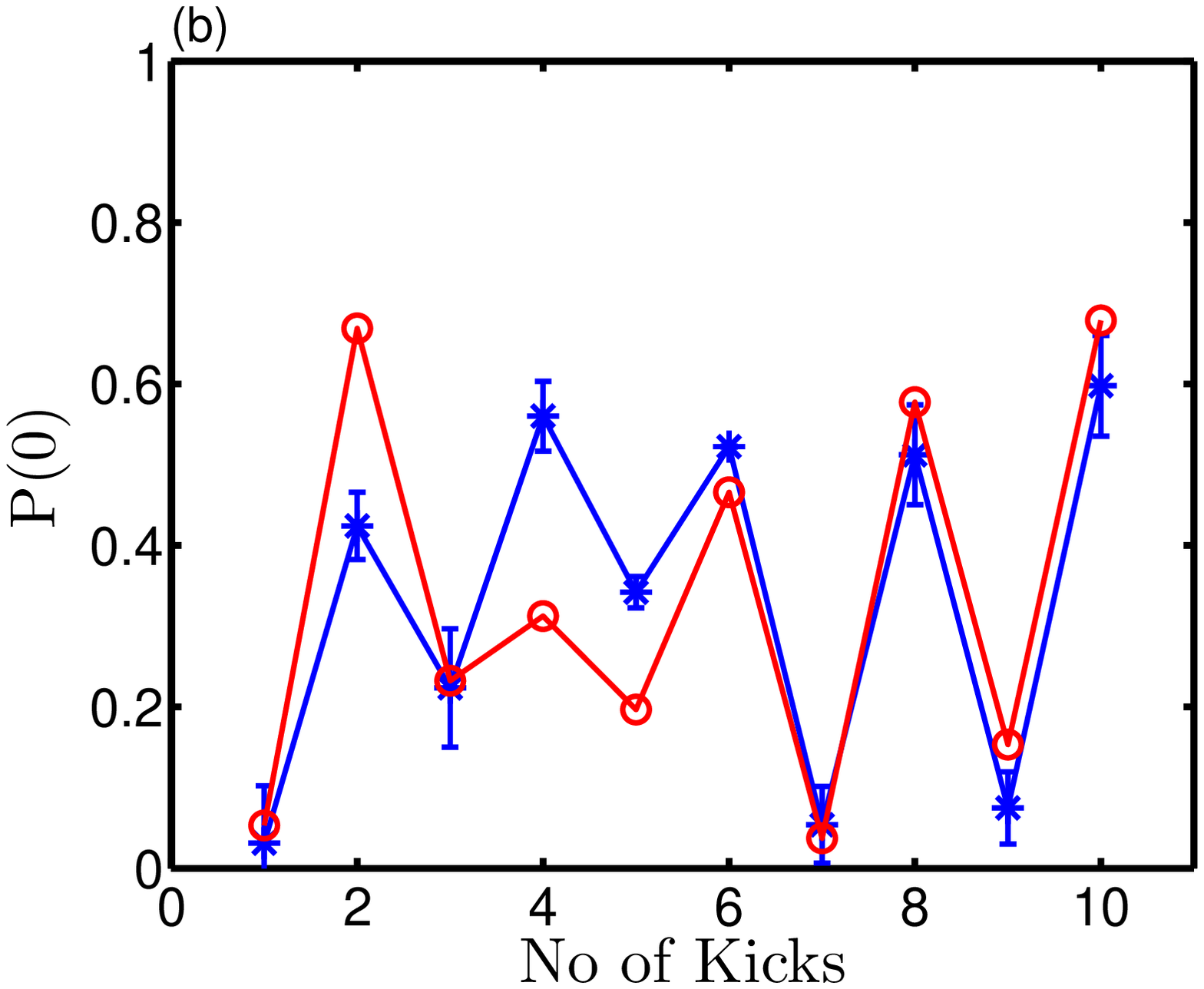}
\caption{ (color online) The normalized height of the zero-momentum
peak P(0), as obtained from the numerical simulation (red circles)
and from the experiment (blue stars). Parameters are $\epsilon=1$
and $\phi_{d}\sim 2$ (a) and $\phi_{d}\sim 3$ (b).
\label{fig:sequence}}
\end{figure}

The BEC we produce in the experiment has a certain finite momentum
width. In order to take into account this initial momentum width in
the simulation, we divide the initial BEC distribution into a number
of components and run the simulation for a range of initial momenta.
We sum over the thus obtained momentum distributions, weighted by
the initial BEC distribution. To account for the limited
experimental resolution, we finally perform a convolution of the
resultant momentum profile from the simulation with a Gaussian, with
a width which is given by the experimental resolution. We can then
determine the height of the momentum peaks obtained corresponding to
various diffraction orders. In both the simulation and the
experiment, the heights of the diffraction orders is summed. We show
the height of the zero momentum peak P(0), which corresponds to the
number of atoms left with in the resolution of the experiment
relative to this sum in Fig.~\ref{fig:sequence}, going through the
kick sequence for two different kick strengths. The error bars in
the experiment are determined by multiple runs of the experiment.

\begin{figure}[h]
\includegraphics[width=\columnwidth]{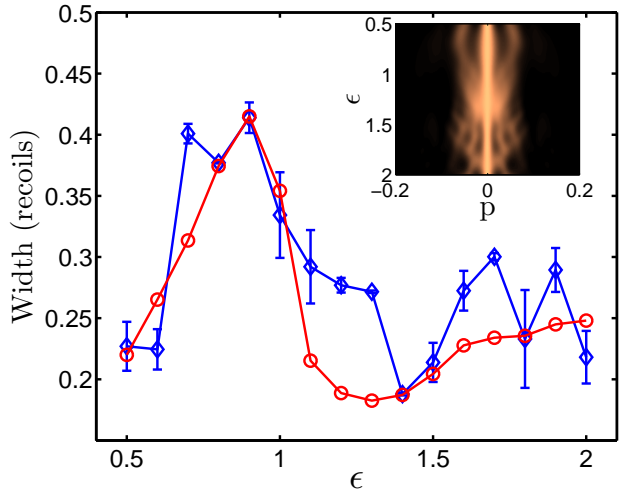}
\caption{\label{fig:width}  (color online) Width of the zero
momentum peak from experiment (blue diamonds) and simulation (red
circles) for a range of $\epsilon$ values.  Also shown (inset) is a
theoretical simulation of the central part of the momentum
distribution for the same range of $\epsilon$ values with no
convolution applied. Other parameters are N=10 and $\phi_{d}\sim
2$.}
\end{figure}

We observe good agreement between the simulation and the experiment
in both cases. The normalized height is small after the first kick,
as the probability is distributed across a number of momentum
states. The height increases after the second kick, as the
probability density for higher momentum states is less, which is
clear from the absorption picture in Fig. \ref{fig:lastpeak}. After
the complete kick sequence, constructive interference of the wave
function components leads to a strong peak at zero momentum. It
should be noted that a maximum height does not necessarily
correspond to the greatest number of ``cold'' atoms remaining, but
rather to the most atoms that appear ''cold'' within the
experimental resolution remaining. From the comparison between the
simulation and experimental results, we believe we have a good
understanding of the parameters in the experiment, and we conclude
that ``Loschmidt'' time reversal has been realized.

In Fig.\ref{fig:width} we plot the width (FWHM) in recoils of the zero momentum peak as a function of $\epsilon$. The error bars shown are obtained by running the experiment a number of times.
Initially we get some higher values of widths for the central peak, which starts to decrease around $\epsilon=1$ and gets narrower at $\epsilon \simeq$=1.3. It should be noted that there are
fewer atoms where the peak is narrower. The results are more clear as seen in the theoretical simulation with no convolution applied in the inset of Fig.\ref{fig:width}. The return probability
for the atoms is equal to 1 (in the simulation). As shown, the final width of the momentum distribution changes for different $\epsilon$. The side-lobes appearing for $\epsilon$ values between
.5 and 1, could be the evidence for the large widths obtained in this range. The lobes then starts to disappear and vanishes at $\epsilon \simeq$ 1.3, which can be seen as a much narrower width
of the momentum distribution for the corresponding value of $\epsilon$. The full structure appearing in the theory plot is difficult to resolve in the experiment because of limited experimental
resolution, but the side-lobes can be observed in Fig. \ref{fig:lastpeak}. The experimentally determined widths, however, follow the general trend of those from simulation after the convolution.

In conclusion, we have experimentally observed evidence for the time
reversal of atomic matter waves in the ultra cold regime. As the
time reversal only works for a narrow range of initial momenta, it
shows a narrowing of the momentum distribution. We have shown this
for a range of parameters in our experiment. In Future, it would be
interesting to add a potential and see the effects of interactions
between atoms as indicated by \cite{Jmartin} on the phenomenon of
time reversibility in quantum chaos.

{\em Acknowledgements:} The authors acknowledge the University of Auckland Research Fund and the Higher Eduction Commission for financial support. The authors would like to thank David Wardle
for fruitful discussions.


\end{document}